\documentclass[11pt]{article}
\usepackage{flafter,amsmath,amssymb,latexsym,psfrag,graphicx,color,bm}
\usepackage[margin=2.5cm]{geometry}
\usepackage{mathtools}
\usepackage[numbers,sort&compress]{natbib}
\usepackage[
linktocpage=true,
colorlinks=true,
linkcolor=blue,
citecolor=blue,
urlcolor=blue
]{hyperref}
\newtheorem{theorem}{Theorem}[section]

\newtheorem{lemma}[theorem]{Lemma}

\allowdisplaybreaks

\makeatletter

\@addtoreset{figure}{section}
\makeatother

\begin{document}
\setlength\arraycolsep{2pt}
\date{}

\title{The factorization method for recovering cavities in a heat conductor}
\author{Jun Guo$^1$, \; Gen Nakamura$^{2}$, \; Haibing Wang$^3$
\\$^1$School of Mathematics and Statistics, South-Central University for Nationalities
\\Wuhan 430074, P.R. China
\\E-mail: hssxgj@126.com
\\$^2$Department of Mathematics, Hokkaido University, Sapporo 060-0810, Japan\qquad\quad
\\E-mail: nakamuragenn@gmail.com
\\$^3$School of Mathematics, Southeast University, Nanjing 210096, P.R. China\qquad\;\;\;\,
\\E-mail: hbwang@seu.edu.cn
}

\maketitle
\begin{abstract}

In this paper, we develop a factorization method to reconstruct cavities in a heat conductor by knowing the Neumann-to-Dirichlet map at the boundary of this conductor. The factorization method is a very well known reconstruction method for inverse problems described by self-adjoint elliptic equations. It enables us to reconstruct the boundaries of unknown targets such as cavities and inclusions. This method was successfully applied for examples to the inverse scattering problem by obstacles for the acoustic wave and the electrical impedance tomography problem. The forward model equations for these problems are the Helmholtz equation and the conductivity equation which are both self-adjoint equations. On the other hand, the heat equation is a typical non-selfadjoint equation. This paper is giving the first attempt to study the factorization method for inverse problems whose forward model equations are non-selfadjoint. We emphasize that an auxiliary operator introduced in this paper is the key which makes this attempt successful. This operator connects the back projection operators for the forward and backward heat equations.



\end{abstract}

\section{Introduction}\label{introduction}
\setcounter{equation}{0}

Let $\Omega\subset \mathbb{R}^n\, (n=2,\,3)$ and $D\Subset\Omega$ be domains with $C^2$ boundaries
$\partial\Omega$ and $\partial D$, respectively. Physically, we consider $\Omega$ as a thermal conductor and $D$ as a cavity embedded in this conductor. Denote by $\nu$ the unit normal vector on the boundary $\partial\Omega$ and $\partial D$. The direction of $\nu$ is always exterior to the domain. That is, for example, $\nu$ on $\partial\Omega$ has the direction pointing into the exterior of $\Omega$. Let the  thermal conductivity of the conductor be $1$ and the initial temperature be zero for simplicity. Inject a heat flux $f$ at the boundary $\partial\Omega$ over the time interval $(0,\,T)$. Then the temperature distribution $u$ in $\Omega\setminus\overline D$ over the time interval $(0,\,T)$ is given as the solution to the following initial boundary value problem:
\begin{equation} \label{1.1}
\begin{cases}
\partial_t u-\Delta u= 0 &  \mathrm{in}\; (\Omega\setminus\overline D)\times (0,\,T),\\
\partial_\nu u=f & \mathrm{on} \; \partial\Omega\times(0,\,T),\\
\partial_\nu u=0 & \mathrm{on} \; \partial D\times(0,\,T),\\
u=0 & \mathrm{at} \; t=0.
\end{cases}
\end{equation}
Since the cavity $D$ is a thermal insulating body which traps heat, the existence of a cavity gives an effect to the temperature distribution $u$ in $\Omega\setminus\overline D$ over the time interval $(0,\,T)$. As a consequence, it also gives an effect to the temperature distribution on the boundary $\partial\Omega$ over the time interval $(0,\,T)$. One can easily speculate that by the existence of a cavity, the temperature distribution on $\partial\Omega$ over the time interval $(0,\,T)$ will increase and it becomes relatively higher on the portion of $\partial\Omega$ where it is closer to $\partial D$. Thermal imaging techniques used in non-destructive testing to identify cavities are based on such a physical phenomenon; see, for example, \cite{CantwellMorton1992, Ibarra-Castanedo2009} and the references therein. One of the thermal imaging techniques called the active thermography is to inject a heat flux $\partial_\nu u\big|_{\partial\Omega\times(0,\,T)}$ at $\partial\Omega$ over the time interval $(0,\,T)$ by a flash lamp or a heater and measure the corresponding temperature distribution $u\big|_{\partial\Omega\times(0,\,T)}$ by using an infrared light camera. This measurement is very quick so that it can be repeated many times. Hence, we mathematically idealize these measurements to consider the so-called Neumann-to-Dirichlet (ND) map $\Lambda_D$ which maps the injected heat flux at $\partial\Omega$ over $(0,\,T)$ to the corresponding temperature distribution at $\partial\Omega$ over $(0,\,T)$. A more precise definition of $\Lambda_D$ will be given later in Section \ref{math model}.  Now we can state the following inverse problem.

\medskip
\noindent
{\bf Inverse problem:} Suppose there is an unknown cavity $D$ inside the heat conductor $\Omega$. Then identify $D$ from the ND map $\Lambda_D$. We are particularly interested in reconstructing $D$ from $\Lambda_D$.

\medskip
The initial boundary value problem \eqref{1.1} which was used to describe our measurement
is called the forward problem in relation with our inverse problem. Its solvability has been already studied in the monograph \cite{Friedman1964} for classical solutions and in \cite{LionsMagenes1972, Costabel1990} for weak solutions. More literatures related to our inverse problems are as follows. The uniqueness and stability can be found in \cite{BryanCaudill1997, DiCristoRondiVessella2006}. The Newton iteration method is used in \cite{BryanCaudill2005, ChapkoKressYoon1998}, and for the non-iterative algorithms, the reader is referred to the probe method \cite{DaidoKangNakamura2007} and the enclosure method \cite{Ikehata2011, IkehataKawashita2010} as well as the references therein. We would like to mention the linear sampling method developed recently in \cite{GuoYanZhou2017, HeckNakamuraWang2012, NakamuraWang2013, NakamuraWang2015, N-W2017, WangLi2018} to reconstruct cavities and inclusions inside a heat conductor. Another kind of sampling-type method, known as the factorization method \cite{Kirsch2008}, is extended to locate the interface in a parabolic-elliptic equation \cite{FruhaufGebauerScherzer2007}.

The factorization method was originated by Kirsch \cite{Kirsch1998} and then widely used in various inverse scattering problems and electrical impedance tomography problems; see, for instance, \cite{Bruhl2001, E-H2019, HankeBruhl2003, HuKirschSini2013, Kirsch2004, Lechleiter2017, Liu2017, Q-Y-Z2017, YangZhangZhang2013}. The equations for these forward problems are self-adjoint elliptic equations. The factorization method is a sampling-type method which gives a criterion whether the sampling points in $\Omega$ are inside the unknown $D\Subset\Omega$ or not. The greatest advantage of this method is that the numerical algorithm of this method is fast and easy to implement. Also, unlike the iterative methods such as the Newton iteration method, it almost does not need any information about the unknown targets. As far as we know, there is not any result on this method for the heat equation except the above mentioned related work \cite{FruhaufGebauerScherzer2007} for a parabolic-elliptic equation. We will show in this paper that it is possible to develop a factorization type method for our inverse problem. For simplicity we will still call this method the factorization method. We will sometimes call the factorization method which has been known so far by {\sl standard factorization method}. We remark that unlike the standard factorization method and that studied in \cite{FruhaufGebauerScherzer2007}, the factorization of our measured data, the ND map, loses the symmetry required by the standard factorization method. However, by introducing an auxiliary operator, we can recover this symmetry.

Concerning the sampling-type method in the time domain, the time domain linear sampling method was developed in \cite{ChenHaddarLechleiterMonk2010, GuoMonkColton2013, HaddarLechleiterMarmorat2014} for inverse scattering problems for the wave equation using near-field data. Further, there is a very recent attempt to adopt the factorization method to recover an obstacle from the far-field measurements of time-dependent causal scattered waves \cite{CakoniHaddarLechleiter2018}.

The rest of this paper is organized as follows. In Section \ref{math model}, we formulate the direct and inverse problems for the heat equation in a conductor including an unknown cavity. By introducing an auxiliary operator which connects the back projection operators for the forward and backward heat equations, we derive in Section \ref{sec3} an appropriate factorization which falls into the abstract framework developed in \cite{Gebauer2006} for the standard factorization method. In Section \ref{sec4}, we show the properties of the operators involved in the factorization. Section \ref{sec_main} is devoted to establishing the factorization method based on the above appropriate factorization. Finally, we give some concluding remarks and discussions in Section \ref{conclusions}.

\section{The precise formulation of the inverse problem}\label{math model}
\setcounter{equation}{0}

In this section, we introduce some anisotropic Sobolev spaces in which we formulate our inverse problem, and then give a detailed description of the inverse problem.

For $r,\,s\geq 0$, we define
$$H^{r,s}(\mathbb{R}^n\times \mathbb{R}):=L^2(\mathbb{R};\,H^{r}(\mathbb{R}^n))\cap H^s(\mathbb{R};\,L^{2}(R^n)).$$
For $r,\,s< 0$, we define by duality
$$H^{r,s}(\mathbb{R}^n\times \mathbb{R}):=(H^{-r,-s}(\mathbb{R}^n\times \mathbb{R}))^\prime.$$
For any bounded domain $E\subset \mathbb{R}^n$, we denote by $H^{r,s}(E_T)$ the space of restrictions of elements in $H^{r,s}(\mathbb{R}^n\times \mathbb{R})$ to $E_T:=E\times (0,\,T)$. The space $H^{r,s}((\partial E)_T)$ are defined analogously when the boundary $\partial E$ of $E$ is a manifold with certain regularity. We also use the following subspaces defined by
$$\widetilde{H}^{1,\frac{1}{2}}(E_T):=\{u\in H^{1,\frac{1}{2}}(E\times(-\infty,\,T))|\,u(x,\,t)=0 ~\mbox{for}~ t< 0\}$$ and 
$$\widehat{H}^{1,\frac{1}{2}}(E_T):=\{u\in H^{1,\frac{1}{2}}(E\times(0,\,+\infty))|\,u(x,\,t)=0 ~\mbox{for}~ t> T\}.$$

By Corollary 3.17 of \cite{Costabel1990}, for any given $f\in H^{-\frac{1}{2},-\frac{1}{4}}((\partial\Omega)_T)$, the forward problem (\ref{1.1}) is well-posed. That is, it has a unique solution $u^f\in \widetilde{H}^{1,\frac{1}{2}}((\Omega\setminus\overline D)_T)$ and it depends continuously on $f$. Also, by using Lemma 2.4 of \cite{Costabel1990} about the continuous Dirichlet trace, we define the ND map more precisely as follows:
\begin{equation}\label{2.1}
\Lambda_D:\, H^{-\frac{1}{2},-\frac{1}{4}}((\partial\Omega)_T)\to {H}^{\frac{1}{2},\frac{1}{4}}((\partial\Omega)_T),\quad f\mapsto u^f|_{(\partial\Omega)_T}.
\end{equation}
Due to the well-posedness of the forward problem and the continuous Dirichlet trace, $\Lambda_D$ is a bounded operator.
Hereafter, whenever we say an operator is bounded means that it is a linear and bounded operator. If $D=\emptyset$, we denote $\Lambda_D$ as $\Lambda_\emptyset$ which is also a bounded operator. Further, let
$$\Gamma(x,\,t;\,y,\,s):=\begin{cases}\displaystyle\frac{1}{[4\pi (t-s)]^{n/2}} \exp\left(-\frac{|x-y|^2}{4(t-s)}\right), & t>s,\\
&\\
\displaystyle 0, &  t\leq s \end{cases}$$
be the fundamental solution of the heat operator $\partial_t-\Delta$ and $G(x,\,t;\,y,\,s)$ be the Green function of the heat operator in $\Omega_T$ with Neumann boundary condition on $(\partial\Omega)_T$. Here $(y,\,s)\in\Omega_T$ denotes the location of singularity for $\Gamma(x,\,t;\,y,\,s)$ and $G(x,\,t;\,y,\,s)$.

If we try to follow the standard factorization method, we will resort to the ND map equation
\begin{equation}\label{2.2}
(\Lambda_D-\Lambda_\emptyset)g=G(\cdot,\,\cdot\,;\,y,\,s)  \quad \textrm{on}\;(\partial\Omega)_T
\end{equation}
for each sampling point $(y,\,s)\in\Omega_T$, and factorize the bounded operator $\Lambda_D-\Lambda_\emptyset$ as 
\begin{equation}\label{2.3}
\Lambda_D-\Lambda_\emptyset=LFL^\prime
\end{equation} 
with a back projection operator $L$ which is called the {\sl virtual measurement operator}, its dual operator $L^\prime$ and a coercive operator $F$. Furthermore, we have to characterize the range of $L$ in terms of $\Lambda_D-\Lambda_\emptyset$; see, for instance, \cite{Bruhl2001, FruhaufGebauerScherzer2007, Gebauer2006} for the details. However, for our inverse problem, we cannot have (\ref{2.3}), since the dual operator $L^\prime$ is associated with the backward heat equation rather than the forward heat equation. We will show how to overcome this difficulty in the next section.

\section{A modified factorization of the ND map}\label{sec3}
\setcounter{equation}{0}

In this section, we modify the factorization \eqref{2.3} of $\Lambda_D-\Lambda_\emptyset$ used in the standard factorization method so that we can put our inverse problem into the abstract framework of the factorization method developed in \cite{Gebauer2006}.

We start by defining our virtual measurement operator $L$ by
\begin{equation}\label{g1}
L:H^{-\frac{1}{2},-\frac{1}{4}}((\partial\Omega)_T)\to {H}^{\frac{1}{2},\frac{1}{4}}((\partial D)_T), ~~ \varphi\mapsto w|_{(\partial D)_T},
\end{equation}
where $w\in \widetilde{H}^{1,\frac{1}{2}}((\Omega\setminus\overline D)_T)$ solves
\begin{equation} \label{3.1}
\begin{cases}
\partial_t w-\Delta w= 0 &  \mathrm{in}\; (\Omega\setminus\overline D)_T,\\
\partial_\nu w=\varphi & \mathrm{on} \; (\partial\Omega)_T,\\
\partial_\nu w=0 & \mathrm{on} \; (\partial D)_T,\\
w=0 & \mathrm{at} \; t=0
\end{cases}
\end{equation}
for $\varphi\in H^{-\frac{1}{2},-\frac{1}{4}}((\partial\Omega)_T)$. Likewise the forward problem (\ref{1.1}), the problem \eqref{3.1} is also well-posed, and hence $L$ is a bounded operator. In order to define a coercive operator $F$ to factorize $\Lambda_D-\Lambda_\emptyset$, we consider the following auxiliary problem:
\begin{equation} \label{3.2}
\begin{cases}
\partial_t v-\Delta v= 0 &  \mathrm{in}\; (\Omega\setminus\overline D)_T,\\
\partial_t v-\Delta v= 0 &  \mathrm{in}\; D_T,\\
\partial_\nu v=0 & \mathrm{on} \; (\partial\Omega)_T,\\
(\partial_\nu v)_{+}-(\partial_\nu v)_{-}=0 & \mathrm{on} \; (\partial D)_T,\\
v_{+}-v_{-}=\psi & \mathrm{on} \; (\partial D)_T,\\
v=0 & \mathrm{at} \; t=0
\end{cases}
\end{equation}
for $\psi\in {H}^{\frac{1}{2},\frac{1}{4}}((\partial D)_T)$. Here the subscripts ``$+$" and ``$-$" denote the traces on $\partial D$ from the exterior and interior of $D$, respectively. By using the mapping properties of the double-layer potential $K_1$ (see Proposition 3.3 and Theorem 3.4 of \cite{Costabel1990}), we can reduce \eqref{3.2} to the following problem for $\tilde v:=v-z$ with the double-layer potential $z:=K_1[\psi]$ with density $\psi$:
\begin{equation}\label{eq for tilde v}
\begin{cases}
\partial_t\tilde v-\Delta\tilde v=0 & \mathrm{in}\;\Omega_T,\\
\partial_\nu\tilde v=-\partial_\nu z\in H^{-\frac{1}{2},-\frac{1}{4}}((\partial\Omega)_T) & \mathrm{on}\;(\partial\Omega)_T,\\
\tilde v=0 & \mathrm{at}\;t=0.
\end{cases}
\end{equation}
Then using the well-posedness of this problem, there exists a unique solution $v\in \widetilde{H}^{1,\frac{1}{2}}((\Omega\setminus\overline D)_T)\cap \widetilde{H}^{1,\frac{1}{2}}(D_T)$ to (\ref{3.2}) which depends continuously on $\psi$. Based on this observation and the continuity of the Neumann trace (\cite[Proposition 2.18]{Costabel1990}), we can define the bounded operator
\begin{equation}\label{g2}
F:H^{\frac{1}{2},\frac{1}{4}}((\partial D)_T)\to {H}^{-\frac{1}{2},-\frac{1}{4}}((\partial D)_T), \quad \psi\mapsto (\partial_\nu v)_{+}|_{(\partial D)_T}.
\end{equation}

Let $\psi=w_{+}|_{(\partial D)_T}$ and define $u_0$ by
\begin{equation}\label{u_0}
u_0:=\begin{cases}
-v & \mathrm{in}\; D_T,\\
w-v & \mathrm{in}\; (\Omega\setminus\overline D)_T.
\end{cases}
\end{equation}
Then it is easy to verify that $u_0$ satisfies
\begin{equation} \label{3.3}
\begin{cases}
\partial_t u_0-\Delta u_0= 0 &  \mathrm{in}\; \Omega_T,\\
\partial_\nu u_0=\varphi & \mathrm{on} \; (\partial\Omega)_T,\\
u_0=0 & \mathrm{at} \; t=0.
\end{cases}
\end{equation}
From the definitions of the ND maps $\Lambda_D$ and $\Lambda_\emptyset$, we have
$$\Lambda_D\varphi=w|_{(\partial\Omega)_T},~~\Lambda_\emptyset\varphi= u_0|_{(\partial\Omega)_T},$$
which immediately yields
\begin{equation}\label{gap eq}
(\Lambda_D-\Lambda_\emptyset)\varphi=(w-u_0)|_{(\partial\Omega)_T} =v|_{(\partial\Omega)_T}.
\end{equation}

Now we define the bounded operator $\widehat{L}$ by
$$\widehat{L}:H^{-\frac{1}{2},-\frac{1}{4}}((\partial D)_T)\to {H}^{\frac{1}{2},\frac{1}{4}}((\partial\Omega)_T), \quad \theta\mapsto v_0|_{(\partial\Omega)_T},
$$
where $v_0\in \widetilde{H}^{1,\frac{1}{2}}((\Omega\setminus\overline D)_T)$ solves
\begin{equation} \label{3.4}
\begin{cases}
\partial_t v_0-\Delta v_0= 0 &  \mathrm{in}\; (\Omega\setminus\overline D)_T,\\
\partial_\nu v_0=0 & \mathrm{in} \; (\partial\Omega)_T,\\
\partial_\nu v_0=\theta & \mathrm{on} \; (\partial D)_T,\\
v_0=0 & \mbox{at} \; t=0
\end{cases}
\end{equation}
for $\theta\in H^{-\frac{1}{2},-\frac{1}{4}}((\partial D)_T)$. 
The boundedness of $\widehat L$ follows from the well-posedness of the problem \eqref{3.4} and the continuity of the Dirichlet trace.
From the definition of $L$, we clearly have
\begin{equation}\label{equality for L}
    L\varphi=w_+\big|_{(\partial D)_T}.
\end{equation}
Since $v_+-v_-=w_+$ on $(\partial D)_T$ from \eqref{u_0}, the definition of $F$ yields
\begin{equation}\label{equality for F}
F(w_+\big|_{(\partial D)_T})=(\partial_\nu v)_+\big|_{(\partial D)_T}.
\end{equation}
Hence, from the definition of $\widehat L$, \eqref{equality for L} and \eqref{equality for F}, we have
\begin{equation}\label{factoring}
v\big|_{(\partial\Omega)_T}=\widehat{L}((\partial_\nu v)_{+}|_{(\partial D)_T}) =\widehat{L}(F(w_{+}|_{(\partial D)_T}))=\widehat{L}(F(L\varphi)).
\end{equation}
Therefore combining \eqref{gap eq} and \eqref{factoring}, we arrive at
\begin{equation} \label{3.5}
\Lambda_D-\Lambda_\emptyset=\widehat{L}FL.
\end{equation}
The factorization (\ref{3.5}) is different from (\ref{2.3}) and it does not have the symmetry. In the following, we modify (\ref{3.5}) to obtain a symmetric factorization formula. To this end, we first compute the dual operator $L^\prime$ of $L$. Here the dual is considered with respect to the continuous extension of the real $L^2((\partial D)_T)$-inner product. 

We begin the argument by considering the following initial boundary value problem:
\begin{equation} \label{3.6}
\begin{cases}
\partial_t \widetilde{w}+\Delta \widetilde{w}= 0 &  \mathrm{in}\; (\Omega\setminus\overline D)_T,\\
\partial_\nu \widetilde{w}=0 & \mathrm{on} \; (\partial\Omega)_T,\\
\partial_\nu \widetilde{w}=\theta & \mbox{on} \; (\partial D)_T,\\
\widetilde{w}=0 & \mbox{at} \; t=T
\end{cases}
\end{equation}
for $\theta\in H^{-\frac{1}{2},-\frac{1}{4}}((\partial D)_T)$. Likewise the  well-posedness of our forward problem, it is known that \eqref{3.6} is well-posed. Hence, the unique solution  $\widetilde{w}\in \widehat{H}^{1,\frac{1}{2}}((\Omega\setminus\overline D)_T)$ to \eqref{3.6} exists and depends continuously on $\theta\in H^{-\frac{1}{2},-\frac{1}{4}}((\partial D)_T)$.

Let $w$ and $\widetilde w$ be the solutions to \eqref{3.1} and \eqref{3.6}, respectively. Then by Green's formula, we have
\begin{eqnarray*}
\langle\varphi,\,L^\prime\theta\rangle&=&\langle\theta,\, L\varphi\rangle=\langle \partial_\nu \widetilde{w}|_{(\partial D)_T}, \,w|_{(\partial D)_T}\rangle
=\int_{(\partial D)_T}\partial_\nu\widetilde{w}w\, dsdt \\
&=&\int_{(\partial\Omega)_T}(\partial_\nu\widetilde{w}w-\partial_\nu{w}\widetilde {w})\,dsdt+\int_{(\partial D)_T}\partial_\nu{w}\widetilde {w}\, dsdt\\
&&-\int_{(\Omega\setminus\overline D)_T}(\Delta\widetilde{w}w-\Delta w\widetilde{w})\,dxdt\\
&=&-\int_{(\partial\Omega)_T}\partial_\nu{w}\widetilde {w}\,dsdt+ \int_{(\Omega\setminus\overline D)_T}(\partial_t\widetilde{w}w+\partial_t w\widetilde{w})\,dxdt\\
&=&-\langle \varphi, \widetilde{w}|_{(\partial\Omega)_T}\rangle +\int_{\Omega\setminus\overline D}[\widetilde{w}(T)w(T)-\widetilde{w}(0)w(0)]\,dx\\
&=&-\langle \varphi, \,\widetilde{w}|_{(\partial\Omega)_T}\rangle,
\end{eqnarray*}
where the notation $\langle\cdot\,,\,\cdot\rangle$ denotes the continuous extension of the real
$L^2((\partial D)_T)$-inner product.
This leads to
\begin{equation}\label{g3}
L^\prime:\,H^{-\frac{1}{2},-\frac{1}{4}}((\partial D)_T)\to {H}^{\frac{1}{2},\frac{1}{4}}((\partial\Omega)_T), \quad \theta\mapsto -\widetilde{w}|_{(\partial\Omega)_T},
\end{equation}
which is bounded due to the well-posedness of \eqref{3.6}.
For the solutions $v_0$ of the problem (\ref{3.4}) and $\widetilde{w}$ to the problem (\ref{3.6}) with the same boundary data $\theta$, we define an auxiliary operator $Q$ by
$$Q:{H}^{\frac{1}{2},\frac{1}{4}}((\partial\Omega)_T)\to {H}^{\frac{1}{2},\frac{1}{4}}((\partial\Omega)_T), \quad v_0|_{(\partial\Omega)_T}\mapsto -\widetilde{w}|_{(\partial\Omega)_T}.$$
Here note that the map $$M:\,H^{-\frac{1}{2},-\frac{1}{4}}((\partial D)_T)\to H^{\frac{1}{2},\frac{1}{4}}((\partial \Omega)_T), \quad \theta\mapsto v_0\big|_{(\partial\Omega)_T}$$
with the solution $v_0$ to \eqref{3.4} has a dense range, which can be shown by an argument similar to the one given in Lemma \ref{Le3.2} for the operators $L$ and $L^\prime$. Then by $\widetilde w(x,\,t)=v_0(x,\,T-t),\,(x,\,t)\in (\Omega\setminus\overline D)_T$, the operator $Q$ can be extended continuously to a bounded invertible operator. We use the same notation $Q$ to denote the extended one. Hence, by the definitions of $\widehat L,\,L^\prime$ and $Q$, we have $\widehat{L}=Q^{-1}L^\prime$. Combining this with (\ref{3.5}), we arrive at the following {\sl modified factorization}
\begin{equation} \label{3.7}
Q(\Lambda_D-\Lambda_\emptyset)=L^\prime FL.
\end{equation}
Also, if we know $\Lambda_D-\Lambda_\emptyset$, we can also know $Q(\Lambda_D-\Lambda_\emptyset)$. Indeed, let $w$ and $u_0$ be solutions of \eqref{3.1} and \eqref{3.3}, respectively. Then since we know $\Lambda_D-\Lambda_{\emptyset}$, we do know $v_0\big|_{(\partial\Omega)_T}$ with $v_0:=w-u_0$. Further, since $v_0$ satisfies \eqref{3.4} with $\theta=-\partial_\nu u_0\big|_{(\partial D)_T}$, we also know $Q(v_0\big|_{(\partial\Omega)_T})$ from the definition of $Q$.
Hence, we can say here that this modified factorization is an acceptable one.

\section{Properties of the operators $L$, $L^\prime$ and $F$}\label{sec4}
\setcounter{equation}{0}
In this section, we investigate the properties of the operators $L,\,L^\prime$ and $F$. To begin with, we first give the following well known lemma (see \cite[Lemma 3.12]{Costabel1990}), which will be used in the proof of Lemma \ref{Le3.1}.

\begin{lemma}\label{Leg1}
Assume $X$ is a real Hilbert space with its dual space $X^\prime$. Let $A:\,X\to X^\prime$ be a bounded operator and $B: \,X\to X^\prime$ be a compact operator. If $A$ and $B$ satisfy
\begin{equation}\label{poitivity of A+B}
\langle(A+B)x,\,x\rangle\geq c_1\|x\|_{X}^2 \quad \textrm{for all}~x\in X
\end{equation}
and
\begin{equation}\label{positive A}
\langle Ax,\,x\rangle> 0 \quad \textrm{for all}~x\in X\setminus\{0\}
\end{equation}
with a constant $c_1>0$, there is a constant $c_2>0$ such that
$$\langle Ax,\,x\rangle\geq c_2\|x\|_{X}^2~~\mbox{for all}~x\in X,$$
where we denote by $\langle\cdot \,,\cdot\rangle$ the pairing between
$X^\prime$ and $X$. We will abbreviate \eqref{positive A} by $A>0$ and call it positive.
\end{lemma}

\begin{lemma}\label{Le3.1}
The operator $-F$ given by (\ref{g2}) is coercive and injective.
\end{lemma}

{\bf Proof.} By recalling (\ref{3.2}), using Green's formula for $v$ and integration by parts, we have
\begin{eqnarray*}
\langle F\psi,\, \psi\rangle
&=&\int_{(\partial D)_T}\partial_\nu v_{+}(v_{+}-v_{-})\, dsdt \\
&=&\int_{(\partial D)_T}(\partial_\nu v_{+}v_{+}-\partial_\nu v_{-}v_{-})\, dsdt \\
&=&\int_{(\partial\Omega)_T}\partial_\nu{v}{v}\,dsdt-\int_{(\Omega\setminus\overline D)_T}(\Delta v v+\nabla v\cdot\nabla v)\, dxdt\\
&&-\int_{D_T}(\Delta v v+\nabla v\cdot\nabla v)\, dxdt\\
&=&-\int_{(\Omega\setminus\overline D)_T}(\partial_t v v+|\nabla v|^2)\, dxdt
-\int_{D_T}(\partial_t v v+|\nabla v|^2)\, dxdt\\
&=&-\int_{(\Omega\setminus\overline D)_T}\frac{1}{2}\partial_t v^2\, dxdt -\int_{(\Omega\setminus\overline D)_T}|\nabla v|^2\, dxdt
-\int_{D_T}\frac{1}{2}\partial_t v^2\, dxdt-\int_{D_T}|\nabla v|^2\, dxdt\\
&=&-\int_{\Omega\setminus\overline D}\frac{1}{2}v^2(T)\, dx
-\int_{(\Omega\setminus\overline D)_T}|\nabla v|^2\, dxdt
-\int_{D}\frac{1}{2}v^2(T)\,dx-\int_{D_T}|\nabla v|^2\, dxdt.\\
&=&-\int_{\Omega\setminus\overline D}\frac{1}{2}v^2(T)dx-\int_{D}\frac{1}{2}v^2(T)\, dx
-\|v\|^2_{H^{1,0}((\Omega\setminus\overline D)_T)}-\|v\|^2_{H^{1,0}(D_T)}\\ &&+\int_{(\Omega\setminus\overline D)_T}|v|^2\, dxdt
+\int_{D_T}|v|^2\, dxdt
\end{eqnarray*}
for any $\psi\in H^{\frac{1}{2},\frac{1}{4}}((\partial D)_T)$, where
$v\in \widetilde H^{1,\frac{1}{2}}((\Omega\setminus\overline D)_T)\oplus\widetilde H^{1,\frac{1}{2}}(D_T)$ is the solution of \eqref{3.2} with the data $\psi$.
Here except the last two terms, the signs of the other terms are minus. We next show that the sum of the last two terms 
can be expressed as
\begin{equation}\label{sesquilinear form with B}
\int_{(\Omega\setminus\overline D)_T}|v|^2\,dxdt+\int_{D_T}|v|^2\,dxdt=\langle B\psi,\, \psi\rangle \end{equation}
with a compact operator $B:\,H^{\frac{1}{2},\frac{1}{4}}((\partial D)_T)\to {H}^{-\frac{1}{2},-\frac{1}{4}}((\partial D)_T)$. 
To see this, let $\widetilde{v}$ be the solution to (\ref{3.2}) with the boundary data $\psi$ replaced by $\zeta$. Then based on the well-posedness of \eqref{3.2}, we can define a bounded and positive sesquilinear form $b:\,H^{\frac{1}{2},\frac{1}{4}}((\partial D)_T)\times H^{\frac{1}{2},\frac{1}{4}}((\partial D)_T)\rightarrow \mathbb{R}$ by
$$b(\psi,\,\zeta)=\int_{(\Omega\setminus\overline D)_T}v\widetilde{v}\,dxdt
+\int_{D_T}v\widetilde{v}\,dxdt \quad \textrm{for any }\psi,\zeta\in H^{\frac{1}{2},\frac{1}{4}}((\partial D)_T).$$
Then by using the Riesz representation theorem, there exists a unique symmetric bounded operator $B:\,H^{\frac{1}{2},\frac{1}{4}}((\partial D)_T)\rightarrow {H}^{-\frac{1}{2},-\frac{1}{4}}((\partial D)_T)$ which satisfies \eqref{sesquilinear form with B}.
The compactness of $B$ follows from the compact embeddings of ${H}^{1,\frac{1}{2}}(D_T)\hookrightarrow L^2(D_T)$ and $H^{1,\frac{1}{2}}((\Omega\setminus\overline D)_T)\hookrightarrow L^2((\Omega\setminus\overline D)_T)$. 
Hence we have
\begin{equation}\label{pre-ineq1}
-\langle (F-B)\psi, \,\psi\rangle\geq \Big(\|v\|^2_{H^{1,0}((\Omega\setminus\overline D)_T)} +\|v\|^2_{H^{1,0}(D_T)}\Big).
\end{equation}

Now by Lemma 2.15 of \cite{Costabel1990}, we have the following equivalences of the norms:
$$
\Vert\cdot\Vert_{{H}^{1,\frac{1}{2}}(D_T)}\sim\Vert\cdot\Vert_{H^{1,0}(D_T)},\,\,\,
\Vert\cdot\Vert_{{H}^{1,\frac{1}{2}}((\Omega\setminus\overline D)_T)}\sim\Vert\cdot\Vert_{H^{1,0}((\Omega\setminus\overline D)_T)}.
$$
Hence we have from \eqref{pre-ineq1} the following inequality
\begin{equation}\label{pre-ineq2}
-\langle (F-B)\psi,\,\psi\rangle\ge c\left(\Vert v\Vert^2_{H^{1,\frac{1}{2}}((\Omega\setminus\overline D)_T)}+\Vert v\Vert^2_{H^{1,\frac{1}{2}}(D_T)}\right)   
\end{equation}
for the solution $v\in \widetilde{H}^{1,\frac{1}{2}}((\Omega\setminus\overline D)_T)\oplus\widetilde{H}^{1,\frac{1}{2}}(D_T)$ of (\ref{3.2}). Then by using the continuity of the Dirichlet trace, we finally arrive at
\begin{equation}\label{last ineq}
\begin{array}{rcl}
-\langle (F-B)\psi, \, \psi\rangle &\geq&
c\Big(\|v_+\|^2_{H^{\frac{1}{2},\frac{1}{4}}((\partial D)_T)} +\|v_-\|^2_{H^{\frac{1}{2},\frac{1}{4}}(\partial D_T)}\Big)\\
&\geq& c\|v_{+}-v_{-}\|^2_{H^{\frac{1}{2},\frac{1}{4}}((\partial D)_T)}\\
&=&c\|\psi\|^2_{H^{\frac{1}{2},\frac{1}{4}}((\partial D)_T)}
\end{array}
\end{equation}
with a general constant $c>0$.
The previous deduction also shows
$$-\langle F\psi,\, \psi\rangle\geq 0.$$

Next we show the operator $F$ is injective. For this purpose, let $F\psi=0$, i.e., $(\partial_\nu v)_+|_{(\partial D)_T}=0$. 
Then from \eqref{3.2}, $v$ satisfies
\begin{equation} \label{3.8}
\begin{cases}
\partial_t v-\Delta v= 0 &  \textrm{in}\; D_T,\\
\partial_\nu v=0 & \textrm{on} \; (\partial D)_T,\\
v=0 & \textrm{at} \; t=0,
\end{cases}
\end{equation}
and thereby $v=0$ in $D_T$. Combining this with the fact that $v$ is the solution to \eqref{3.2}, $v$ satisfies
\begin{equation}\label{3.8 minus}
\begin{cases}
\partial_t v-\Delta v= 0 &  \textrm{in}\; (\Omega\setminus\overline D)_T,\\
\partial_\nu v=0 & \textrm{on}\; (\partial\Omega)_T,\\
\partial_\nu v=0 & \textrm{on} \; (\partial D)_T,\\
v=0 & \textrm{at} \; t=0,
\end{cases}
\end{equation}
and thereby $v=0$ in $(\Omega\setminus\overline D)_T$. So we have
$\psi=v_{+}|_{(\partial D)_T}-v_{-}|_{(\partial D)_T}=0$, and thus the injectivity of $F$ is proved. Consequently, we have $-F>0$. By using Lemma \ref{Leg1}, we obtain the coercivity of $-F$. The proof is complete. \hfill $\Box$

\begin{lemma}\label{Le3.2}
The operators $L$ and $L^\prime$ given by (\ref{g1}) and (\ref{g3}), respectively, are compact, injective and therefore have dense ranges.
\end{lemma}

{\bf Proof.} Let $G^0(\cdot,\,\cdot\,\,;y,\,s)$ be the Green function of the heat operator in the domain $(\mathbb{R}^n\setminus\overline D)_T$ with the Neumann boundary condition on $(\partial D)_T$. Then the solution of the problem (\ref{3.1}) can be expressed uniquely by the single-layer potential
$$w^{\alpha}(x,\,t)=\int_{(\partial\Omega)_T}G^0(x,\,t;\,y,\,s)\alpha(y,\,s)\,ds(y)ds, \quad x\in\Omega\setminus \overline D, ~0< t< T, $$
where the density $\alpha$ belongs to $H^{-\frac{1}{2},-\frac{1}{4}}((\partial\Omega)_T)$. Since $w^\alpha|_{(\partial D)_T}$ is a smooth function on the closure of $(\partial D)_T$, $L$ is compact and so is it for $L^\prime$.

Since the injectivity of $L$ implies the denseness of $L^\prime$, and the same by interchanging $L$ with $L^\prime$. The injectivity properties of $L$ and $L^\prime$ can be proved in the same way, so we only show here the injectivity for $L$. To this end, let $L\varphi=w|_{(\partial D)_T}=0$, where $w\in \widetilde{H}^{1,\frac{1}{2}}((\Omega\setminus\overline D)_T)$ is the solution to \eqref{3.1}. Then the Cauchy data of $w$ on $(\partial D)_T$ are zero. Hence by the unique continuation property ({\sl UCP}) of solutions of the heat equation, we have $w=0$ in $(\Omega\setminus\overline D)_T$, which implies $\varphi=\partial_\nu w\big|_{(\partial\Omega)_T}=0$. 
The proof is complete. \hfill $\Box$

\section{The factorization method}\label{sec_main}
\setcounter{equation}{0}

In this section, we provide the factorization method for our inverse problem. To begin with, we cite the following two results from functional analysis given as lemmas.
\begin{lemma}\label{Le3.3}
Let $H$ be a real Hilbert space. Assume that the bounded operator $A$ on $H$ is a positive and symmetric operator. Then the operator $A$ has a unique positive and symmetric root $A^{1/2}$ which is a bounded operator on $H$. Moreover, $A$ has a decomposition in the form
$$A=(A^{1/2})(A^{1/2})^*,$$
where $A^\ast$ denotes the adjoint of $A$.
If $A$ is bijective and so is $A^{1/2}$. Once again we note that
a bounded operator means that it is a linear operator and bounded
in this paper.
\end{lemma}

This is a very well known fact in functional analysis.
\begin{lemma}\label{Le3.4}
Let $H_1$ and $H_2$ be real Hilbert spaces, and $X$ be an another real Hilbert space. Assume the bounded operators $A_j:\,H_j\to X$, $j=1,2$ satisfy $A_1A_1^*=A_2A_2^*$. Then the ranges of $A_1$ and $A_2$ coincide.
\end{lemma}

This lemma is given as Lemma 3.5 of \cite{Gebauer2006}.

\medskip
Before giving the arguments of this section, let us clarify the notations frequently used in this section. For a real Hilbert space $H$ and its dual space $H^\prime$, we denote by $(\cdot\, ,\,\cdot)$ the inner product in $H$ and by $\langle\cdot\, , \,\cdot\rangle$ the dual pairing on $H^\prime \times H$. They have the relation $\langle l_H u,\,\cdot\rangle = (u,\,\cdot)$ for all $u\in H$ in terms of the isometry $l_H:\,H\to H^\prime$. For a bounded operator $A$ between real Hilbert spaces $H_1$ and $H_2$, the dual operator $A^\prime$ and the adjoint operator $A^*$ satisfy the following relation $A^*=l_{H_1}^{-1}A^\prime l_{H_2}$. Hence, by $l_{H_1^\prime}=l_{H_1}^{-1}$, we have
\begin{equation}\label{W1}
(Al_{H_1^\prime})^*=A^\prime l_{H_2}.
\end{equation}

Set $N:=-Q(\Lambda_D-\Lambda_\emptyset)$ for simplicity. Then (\ref{3.7}) can be written as
\begin{equation} \label{3.11}
N=L^\prime(-F)L.
\end{equation}
We call $N$ the {\sl modified Neumann-to-Diriclet map} (modified ND map). In order to apply Lemmas \ref{Le3.3} and \ref{Le3.4}, we need to consider the symmetrization of (\ref{3.11}). To this end, we set
\begin{eqnarray}
\mathcal{N}=\frac{1}{2}(N+N^\prime),\quad \widetilde{N}=\mathcal{N}l_{{H}^{\frac{1}{2},\frac{1}{4}} ((\partial\Omega)_T)},\label{3.12}\\
\mathcal{F}=\frac{1}{2}(-F-F^\prime),\quad \widetilde{F}=l_{{H}^{\frac{1}{2},\frac{1}{4}} ((\partial D)_T)}^{-1}\mathcal{F}\label{3.13}.
\end{eqnarray}
One observes that $\mathcal{N}=\mathcal{N}^\prime$, $\mathcal{F}=\mathcal{F}^\prime$ and $(L^\prime)^\prime=L$. Further by applying \eqref{W1} with $A=L^\prime$, $H_1=H^{-\frac{1}{2},-\frac{1}{4}}((\partial D)_T)$ and $ H_2=H^{\frac{1}{2},\frac{1}{4}}((\partial\Omega)_T)$, we have
\begin{eqnarray}\label{3.14}
\widetilde{N}&=&L^\prime\mathcal{F}Ll_{{H}^{\frac{1}{2},\frac{1}{4}} ((\partial\Omega)_T)}
=L^\prime l_{{H}^{\frac{1}{2},\frac{1}{4}} ((\partial D)_T)} \widetilde{F}Ll_{{H}^{\frac{1}{2},\frac{1}{4}} ((\partial\Omega)_T)}\nonumber\\[2mm]
&=&L^\prime l_{{H}^{\frac{1}{2},\frac{1}{4}} ((\partial D)_T)} \widetilde{F}(L^\prime l_{{H}^{\frac{1}{2},\frac{1}{4}} ((\partial D)_T)})^*.
\end{eqnarray}
This decomposition is the key to the factorization method. 
Our next task is to show that the range of $L^\prime$ can be characterized by the modified ND map. We start by preparing the following lemma.

\begin{lemma}\label{Le3.5}
(a) The operator $\widetilde{F}:\,{H}^{\frac{1}{2},\frac{1}{4}} ((\partial D)_T)\to {H}^{\frac{1}{2},\frac{1}{4}} ((\partial D)_T)$ is self-adjoint and coercive, and hence surjective.
(b) The operator $\widetilde{N}:\,{H}^{\frac{1}{2},\frac{1}{4}} ((\partial\Omega)_T) \to {H}^{\frac{1}{2},\frac{1}{4}} ((\partial\Omega)_T)$ is self-adjoint and positive.
\end{lemma}

{\bf Proof.} (a) For any $\eta_1,\,\eta_2\in {H}^{\frac{1}{2},\frac{1}{4}} ((\partial D)_T)$, by using the symmetry of $\mathcal{F}$, a direct calculation gives
\begin{eqnarray*}
(\widetilde{F}\eta_1,\,\eta_2)&=&(l_{{H}^{\frac{1}{2},\frac{1}{4}} ((\partial D)_T)}^{-1}\mathcal{F}\eta_1,\,\eta_2)=\langle\mathcal{F}\eta_1,\eta_2\rangle\\[2mm]
&=&\langle\mathcal{F}\eta_2,\,\eta_1\rangle=(l_{{H}^{\frac{1}{2},\frac{1}{4}} ((\partial D)_T)}^{-1}\mathcal{F}\eta_2,\,\eta_1)\\[1mm]
&=&(\eta_1,\widetilde{F}\eta_2),
\end{eqnarray*}
which shows that $\widetilde{F}$ is self-adjoint. Due to the coerciveness of $-F$, we have
\begin{eqnarray*}
(\widetilde{F}\eta_1,\,\eta_1)&=&\langle\mathcal{F}\eta_1,\,\eta_1\rangle
=-\frac{1}{2}\langle F\eta_1,\,\eta_1\rangle-\frac{1}{2}\langle F^\prime\eta_1,\,\eta_1\rangle\\
&=&-\langle F\eta_1,\,\eta_1\rangle\geq c\|\eta_1\|^2_{{H}^{\frac{1}{2},\frac{1}{4}} ((\partial D)_T)}.
\end{eqnarray*}

(b) The proof is quite similar to (a). For $\rho_1,\,\rho_2\in {H}^{\frac{1}{2},\frac{1}{4}}((\partial\Omega)_T)$, we can deduce \begin{eqnarray*}
(\widetilde{N}\rho_1,\,\rho_2)&=&(\mathcal{N}l_{{H}^{\frac{1}{2},\frac{1}{4}}((\partial \Omega)_T)}\rho_1,\,\rho_2)=\langle \mathcal{N}l_{{H}^{\frac{1}{2},\frac{1}{4}}((\partial \Omega)_T)}\rho_1,\,l_{{H}^{\frac{1}{2},\frac{1}{4}}((\partial \Omega)_T)} \rho_2\rangle\\[2mm]
&=&\langle \mathcal{N}l_{{H}^{\frac{1}{2},\frac{1}{4}}((\partial \Omega)_T)} \rho_2,\,l_{{H}^{\frac{1}{2},\frac{1}{4}}((\partial \Omega)_T)}\rho_1\rangle
=(\mathcal{N}l_{{H}^{\frac{1}{2},\frac{1}{4}}((\partial \Omega)_T)}\rho_2,\,\rho_1)\\[1mm]
&=&(\rho_1,\,\widetilde{N}\rho_2),
\end{eqnarray*}
which means that $\widetilde{N}$ is self-adjoint. We have from (\ref{3.14}) together with the coerciveness of the operator $\widetilde{F}$ that
\begin{eqnarray*}
(\widetilde{N}\rho_1,\,\rho_1)&=&(L^\prime l_{{H}^{\frac{1}{2},\frac{1}{4}} ((\partial D)_T)} \widetilde{F}Ll_{{H}^{\frac{1}{2},\frac{1}{4}}((\partial\Omega)_T)}\rho_1,\,\rho_1)\\[2mm]
&=&(\widetilde{F}Ll_{{H}^{\frac{1}{2},\frac{1}{4}}((\partial\Omega)_T)}\rho_1,\, Ll_{{H}^{\frac{1}{2},\frac{1}{4}}((\partial\Omega)_T)}\rho_1)\\[2mm]
&\geq& c\|Ll_{{H}^{\frac{1}{2},\frac{1}{4}}((\partial\Omega)_T)}\rho_1\|^2 _{{H}^{\frac{1}{2},\frac{1}{4}} ((\partial D)_T)}\\[2mm]
&>& 0 \quad\mbox{for}~\rho_1\neq 0.
\end{eqnarray*}
The proof is complete.   \hfill $\Box$

\medskip
Now we are ready to characterize the range of $L^\prime$ by using the modified ND map as follows.
\begin{lemma}\label{Th3.6} The ranges of the operator $\widetilde{N}^{1/2}:\,{H}^{\frac{1}{2},\frac{1}{4}}((\partial\Omega)_T)\to {H}^{\frac{1}{2},\frac{1}{4}}((\partial\Omega)_T)$ and the operator $L^\prime:\,{H}^{-\frac{1}{2},-\frac{1}{4}}((\partial D)_T)
\to {H}^{\frac{1}{2},\frac{1}{4}}((\partial\Omega)_T)$ coincide.
\end{lemma}

{\bf Proof.} From Lemmas \ref{Le3.3} and \ref{Le3.5}, $\widetilde{F}$ and $\widetilde{N}$ have self-adjoint and bijective square roots $\widetilde{F}^{1/2}$ and $\widetilde{N}^{1/2}$, respectively. Then by (\ref{3.14}), we arrive at
$$\widetilde{N}=\widetilde{N}^{1/2}(\widetilde{N}^{1/2})^*=L^\prime l_{{H}^ {\frac{1}{2},\frac{1}{4}} ((\partial D)_T)} \widetilde{F}^{1/2} (\widetilde{F}^{1/2})^*(L^\prime l_{{H}^{\frac{1}{2},\frac{1}{4}} ((\partial D)_T)})^*.
$$
Lemma \ref{Le3.4} yields
\begin{equation}
\label{range}
\mathcal{R}(\widetilde{N}^{1/2})=\mathcal{R}(L^\prime l_{{H}^ {\frac{1}{2},\frac{1}{4}} ((\partial D)_T)} \widetilde{F}^{1/2}).
\end{equation}
Here note that the operator $l_{{H}^ {\frac{1}{2},\frac{1}{4}} ((\partial D)_T)}$ is surjective. Also, from the surjectivity of $\widetilde F$ and $\mathcal{R}(\widetilde F)=\mathcal{R}(\widetilde F^{1/2}\widetilde F^{1/2})\subset\mathcal{R}(\widetilde F^{1/2})$, the operator $\widetilde{F}^{1/2}$ is surjective. Taking these into account, the assertion follows immediately from \eqref{range}.
\hfill $\Box$

\medskip
Next by using the Green function $G^\prime(\cdot,\,\cdot\, ; y,\,s)$ of the backward heat operator $\partial_t+\Delta$ in $\Omega_T$ with Neumann boundary condition and a singularity at $(y,\,s)$ as the test function, we can show that $G^\prime(\cdot,\,\cdot\, ;\, y,\,s)|_{(\partial\Omega)_T}\in \mathcal{R}(L^\prime)$ if and only if the sampling point $(y,\,s)\in D_T$. Hence by applying Picard's theorem (\cite{Kirsch2011}), an indicator function can be set up to determine the cavity $D$ from the modified ND map.

\begin{lemma}\label{Le3.7}
The Green function $G^\prime(\cdot,\,\cdot\, ;\, y,\,s)|_{(\partial\Omega)_T}$ for $(y,\,s)\in \Omega_T$ belongs to the range of $L^\prime$ if and only if $(y,\,s)\in D_T$.
\end{lemma}

{\bf Proof.} We first consider the case $(y,\,s)\in D_T$. Since $G^\prime(\cdot,\,\cdot\, ;\, y,\,s)$ solves (\ref{3.6}) with boundary data $\theta=(\partial_\nu G^\prime(\cdot,\,\cdot\, ;\, y,\,s))|_{(\partial D)_T}$, we can easily see $$L^\prime((\partial_\nu G^\prime(\cdot,\,\cdot\, ;\, y,\,s))|_{(\partial D)_T})=-G^\prime(\cdot,\,\cdot\, ;\, y,\,s)|_{(\partial\Omega)_T}.$$

Next we consider the case $(y,\,s)\in (\Omega\setminus D)_T$. Suppose we have $L^\prime\theta=-G^\prime(\cdot,\,\cdot\, ; \,y,\,s)|_{(\partial\Omega)_T}$ for some $\theta\in {H}^{-\frac{1}{2},-\frac{1}{4}}((\partial D)_T)$. Let $\widetilde{w}$ be the solution to (\ref{3.6}) with this boundary data $\theta$. By Lemma \ref{Le3.2}, $L^\prime$ is injective and hence we have $\widetilde{w}|_{(\partial\Omega)_T}=G^\prime(\cdot
,\,\cdot\, ;\, y,\,s)|_{(\partial\Omega)_T}$. Since $\partial_\nu G^\prime(\cdot,\,\cdot\, ;\, y,\,s)|_{(\partial\Omega)_T}=(\partial_\nu \widetilde{w})|_{(\partial\Omega)_T}=0$, by the UCP for the operator $\partial_t +\Delta$ (\cite{Isakov2006}), we have
$$\widetilde{w}=G^\prime(\cdot,\,\cdot\, ;\, y,\,s) \quad \mathrm{in} ~(\Omega\setminus(\overline D\cup\{y\}))_T,$$
which leads to a contradiction, because $\widetilde{w}$ belongs to $\widehat{H}^{1,\frac{1}{2}}((\Omega\setminus\overline D)_T)$ but $G^\prime(\cdot,\,\cdot\, ;\, y,\,s)$ does not, due to the fact that $G^\prime(\cdot,\,\cdot\, ;\, y,\,s)$ has the same singularity as the fundamental solution of the backward heat operator.
\hfill $\Box$

\medskip
Now observe from (\ref{3.14}) and the compactness of $L^\prime$ that $\widetilde{N}$ is compact. Then $\widetilde{N}^{1/2}$ is also compact by the spectral decomposition. Combining Lemma \ref{Th3.6}, Lemma \ref{Le3.7} and Picard's theorem, we have the following main result.
\begin{theorem}\label{Th3.8}
Let $\{\lambda_n,\,\psi_n\}$ be an eigensystem of the operator $\widetilde{N}$ with normalized eigenfunctions and consider the equation
\begin{equation}\label{3.15}
\widetilde{N}^{1/2}g_{(y,\,s)}=G^\prime(\cdot,\,\cdot\, ;\, y,\,s)|_{(\partial\Omega)_T}\quad \mbox{for}~(y,\,s)\in\Omega_T.
\end{equation}
Then we have
$$(y,\,s)\in D_T \Longleftrightarrow G^\prime(\cdot,\,\cdot\, ;\, y,\,s)|_{(\partial\Omega)_T}\in \mathcal{R}(\widetilde{N}^{1/2}),$$
and consequently
$$
(y,\,s)\in D_T \Longleftrightarrow \sum_{n=1}^{\infty}\frac{|(G^\prime(\cdot,\,\cdot \,;\, y,\,s),\,\psi_n)|^2}{\lambda_n}<\infty.
$$
In other words, the indicator function
\begin{equation}\label{3.16}
W(y,\,s):=\left[\sum_{n=1}^{\infty}\frac{|(G^\prime(\cdot,\,\cdot\, ;\, y,\,s),\,\psi_n)|^2} {\lambda_n}\right]^{-1}=
\left\{
\begin{array}{ll}
c\in (0,\,+\infty), &\,\,(y,\,s)\in D_T,\\
0, &\,\,(y,\,s)\not\in D_T
\end{array}
\right.
\end{equation}
can be used to recover $D$.
\end{theorem}

\section{Conclusions}\label{conclusions}

In this paper, we developed the factorization method for the inverse problem of reconstructing cavities in a thermal conductor from the measurement data $\Lambda_D$. Note that the standard factorization of the operator $\Lambda_D - \Lambda_\emptyset$ does not satisfies the symmetry required by the well-known argument. We introduced an auxiliary operator $Q$ which connects the back projection operators for the forward and backward heat equations, and then deduced a modified factorization of the form $Q(\Lambda_D - \Lambda_\emptyset) = L^\prime F L$. After showing some properties of the operators $L,\,L^\prime$ and $F$, we justified the factorization method for our inverse problem by characterizing the range of the operator $L^\prime$. Our work provides an insight to develop the factorization method for non-selfadjoint governing equations, and can be extended to handle inverse boundary value problems for more general non-selfadjoint models.

\bigskip
{\bf Acknowledgement:} During this study GN was supported by Grant-in-Aid for Scientific Research of the Japan Society for the Promotion of Science (Nos. 15K17555, 15H05740), and HW was supported by National Natural Science Foundation of China (Nos.11671082, 11971104). We greatly appreciate these supports.

\end{document}